# Characterization of low-viscosity electrorheological fluids: Technical Issues and Challenges

Pedro. C. Rijo[1,2] and Francisco J. Galindo-Rosales[2,3]*

[1] CEFT – Transport Phenomena Research Center, Mechanical Engineering Dept., Faculty of Engineering, University of Porto, Rua Dr. Roberto Frias, 4200-465 Porto, Portugal.

[2] AliCE – Associate Laboratory in Chemical Engineering, Faculty of Engineering, University of Porto, Rua Dr. Roberto Frias, 4200-465 Porto, Portugal.

[3] CEFT – Transport Phenomena Research Center, Chemical Engineering Dept., Faculty of Engineering, University of Porto, Rua Dr. Roberto Frias, 4200-465 Porto, Portugal.

* Corresponding author: galindo@fe.up.pt

**Abstract:** The electrorheological (ER) characterization of low-viscosity fluids is paramount for producing micro- and nanoscale products through electrohydrodynamic (EHD) techniques, such as EHD-jet printing, electrospray, and electrospinning. Key properties such as viscosity, surface tension, dielectric properties, electrical conductivity, and relaxation time significantly influence both the quality and properties of the final products and the efficiency of the industrial process. ER characterization is essential for studying the macroscopic effects of the interaction between these physicochemical properties under controlled flow kinematics.

Researchers may face several technical challenges in performing rigorous ER characterization of low-viscosity fluids. This characterization is crucial for formulating inks compatible with the EHD process and for understanding fluid dynamics in EHD processes to ensure stable printing conditions and achieve high-resolution, accurate prints. This work highlights the inherent limitations of current ER cells and proposes methodologies to mitigate their impact on measurement accuracy. Furthermore, we explore the potential of microfluidic technologies to offer innovative solutions for the ER characterization of low-viscosity fluids.

# 1. Introduction

Electrohydrodynamic (EHD) techniques have increased interest from the scientific community since the beginning of the XXI century. It works by applying an electric field to the induce fluid ejection from a conductive nozzle onto a substrate [1]. The needs to produce new micro- and nanoscale products has increased to solve challenges present in medicine, electronics, robotics, and additive manufacturing applications.[2] The main EHD techniques are the EHD-jet printing, electrospinning, and electrospray (**Figure 1 a-c**). These techniques have been an alternative with low production costs to the traditional manufacturing techniques, such as lithography, inkjet printing, screen printing, etc.[1] The quality of the final products made by EHD techniques depends on a set of parameters, which can be divided into two main groups: operating parameters and fluid parameters. [1-3] The operating parameters are the applied voltage, the distance between the needle and the collector plate (working distance), and the flow rate. The fluid parameters are mainly the viscosity, the surface tension, and the electrical properties, which depend strongly on the polymer/particles concentration of the fluid and the solvent's properties. **Table 1** shows the threshold values of some properties mentioned above for each EHD technique.

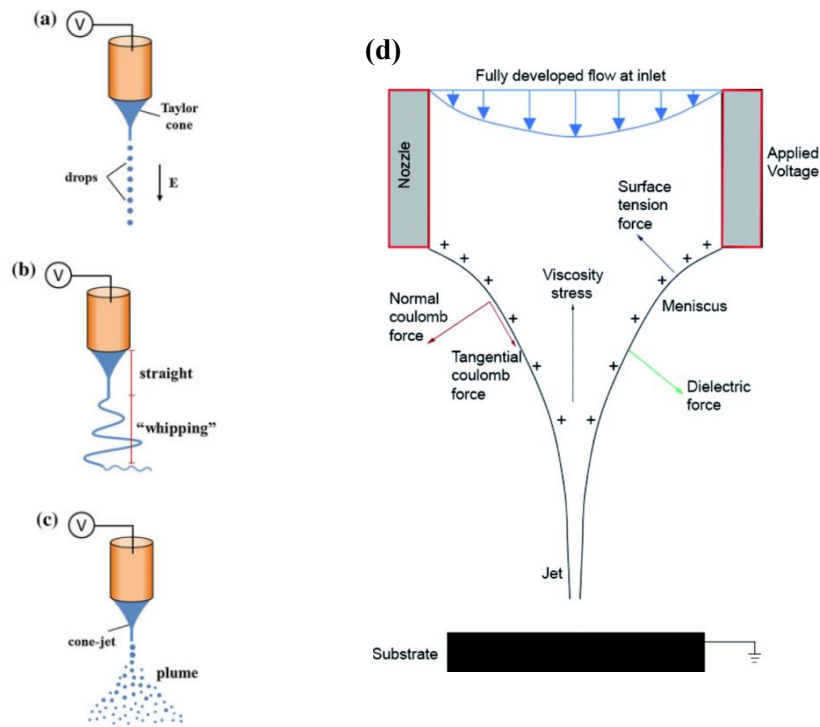

**Figure 1**. Schematic illustration of the EHD jet printing **(a)**, electrospinning **(b)**, and electrospray **(c)**. Schematic representation of the forces acting on a Taylor cone during an EHD process **(d)**. Adapted from [4, 5].

**Table 1** – The threshold values of the main parameters that affects the EHD techniques. Adapted from [4].

|  | **EHD-Jet printing** | **Electrospinning** | **Electrospray** |
|---|---|---|---|
| **Voltage [kV]** | 0.5 - 3 | 1 - 15 | 10 - 30 |
| **Working distance [mm]** | 0.1 - 1 | 10 - 50 | 100 - 250 |
| **Viscosity [mPa·s]** | < 100 | 100 - 10000 | < 50 |
| **Surface Tension [mN/m]** | 20 - 50 | 15 - 64 | < 50 |

Independently of the EHD technique used, the Taylor's cone will be present whenever there is an electric field applied to the fluid. This Taylor's cone formation occurs as the potential difference creates an electric field around the liquid meniscus attached to the nozzle tip and brings the electric charges or ions to the meniscus surface, which promotes the elongation of the meniscus and eventual its breakup in the direction of the applied electric field.[5] Then, once the different forces present in the meniscus (**Figure 1d**) reach an equilibrium, Taylor's cone is formed. The electric forces (Coulomb and dielectric) developed at the meniscus surface are balanced by the surface tension as well as the reverse viscous flow induced in the liquid. Traditionally, the fluid viscosity considered for the dimensionless analysis was measured in the absence of an electric field; however, depending on the electrorheological properties of the working fluid, it is known that the fluid's viscosity can be significantly modified by the presence of an external electric field. Thus, the lack of knowledge of the electrorheological behavior does not allow to know if some technical issues, such as nozzle clogging, are derived from the evaporation of the solvent by Joule's effect due to the electric field or if they are caused by a sudden increase of the fluid viscosity in the presence of electric field. Therefore, the electrorheological properties of the fluids used in EHD techniques need to be carefully studied to determine in order to optimize the EHD printing processes and avoid technical issues.

Electrorheology is a branch of rheology dealing with the study of the flow and deformation of fluids in the presence of electric field. It was originated after Willis M. Wilson discovered the electrorheological effect [6] by which certain liquid suspensions exhibited an sudden increase in viscosity due to the electrical field-induced columnar structures of particles, analogously to the fibration observed in magnetic fluid under the presence of an external magnetic field [7]. This effect occurs when a suspension, colloid or material in the suspended state contains dielectric particles, which will agglomerate or be arranged in chains or columns along the direction of the electric field, resulting in an increase of the apparent viscosity, and even exhibiting yield stress. This effect is also known as positive ER.[8] However, the ER effect can also be negative; in the negative ER effect, the viscosity decreases in the presence of an electric field, and this phenomenon has been far less studied than the positive ER effect [9].

Since the discovery of the (positive) ER effect, many theories have emerged to explain it. In the 20th century, three theories were proposed: fiber theory[7], electric double layer polarization theory [10], and "water bridge" theory.[11] These theories have their limitations: the former theory cannot explain the slow rate fibrillation and the difference between the millisecond response times of the ER effect. The last two theories can only explain the ER effect for fluids having water in their composition. In recent years, four new theories have gained popularity and have been accepted by the scientific community: dielectric theory,[12, 13] the orientation and bonding theory of polar molecules [14-16], the surface polarization saturation model [17] and the saturated orientational polarization model [18]. The last three theories have been used to explain ER fluids show a yield shear stress above 100 kPa, known as giant electrorheological fluids, and have polar molecules in their composition. In addition to the abovementioned theories to explain the ER effect, many rheological models have been developed to characterize the fluid behavior: Bingham model [19], Herschel-Buckley model[19], De Kee-Turcotte model [20], Seo-Seo model [21, 22], and Cho-Choi-Jhon model [23]. Most of these models need to be fitted to find the suitable parameter values for different types of ER fluids and for a specific electric field strength.

During the ER fluid's formulation, it is necessary to consider a several parameters that directly influence the ER effect. The carrier fluid must have a low dielectric properties and electrical conductivity and the dispersed phase, that can be either a solid or a liquid material, must have a high dielectric constant, an appropriate electrical conductivity, density, and relatively stable physicochemical properties [8, 10]. Dielectric inorganics (*e.g.,* metal oxides, silicate materials) [4, 24], conductive organics (*e.g.,* graphene, carbon nanotubes) [4, 24] and polymers (*e.g.,* polyaniline, polypyrrole) [25-32], biopolymers (*e.g.,* cellulose derivatives)[24, 33], and liquid crystals (*e.g.,* cyanobiphenyls) [34-36] are common dispersed materials used to formulate ER fluids [14].  Further, the performance of ER fluids also depends on the volume concentration, size, shape, and geometry of the particles used as dispersed phase [14]. **Figure 2** schematically shows the relative weighting of the available works published in literature for each type of dispersed phase mentioned above.

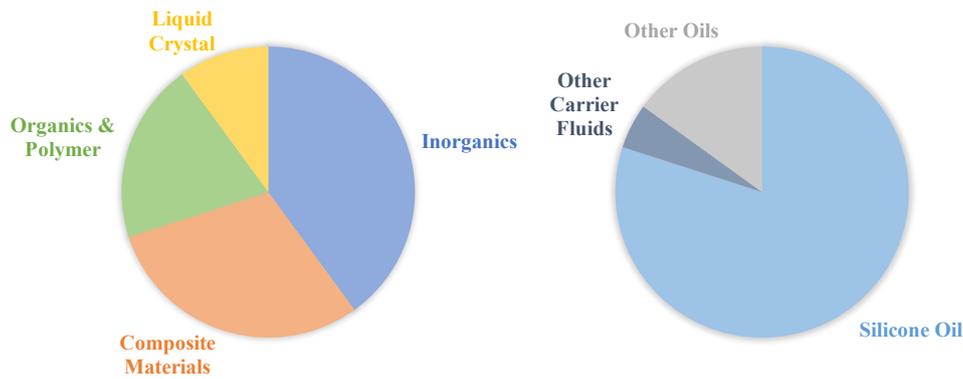

**Figure 2.** Relative weighting for each type of dispersed phase and carrier fluids were used to study the ER effect. Adapted from [24].

According to Kuznetsov *et al.* [24], silicone oil has been the most used solvent to produce ER fluids followed by mineral oil, castor oil, etc. due to their weak dielectric properties and electrical conductivity. The authors found that the minimum viscosity used for silicone oil was 10 mPa·s, and independently of the particle's concentration used in the formulation of the ER fluid, the yield shear stress and the shear viscosity can be higher than 5 Pa and 1 Pa·s, respectively, when an external electric field is applied. Based on this threshold values, the scientific community have been focusing on the formulation of ER fluids that may be used in aerospace, robotics, clutches, shock-absorbers, valves, and quick-response dampers [10, 24, 37], where the ER fluid must have a positive ER effect. When the ER fluids show a negative ER effect, the scientists have been focusing on crude oil transportation and on the food processing, e.g., reducing the fat content in chocolate [10]. When the ER fluids have a yield shear stress or a shear viscosity below the threshold values mentioned above, a few studies are found in literature. In this area of research, M.T. Cidade's group [38] focused on the electrorheological characterization of ER fluids where the dispersed phase is liquid crystal [34-36] and hybrid/composite particles [30, 31] made of polymer and inorganic/organic particles.

**Table 2.** Properties of the most used solvents in EHD processes. Adapted from [4].

| Solvents | Density [g/mL] | Surface Tension [mN/m] | Dielectric Constant [-] |
|---|---|---|---|
| **Chloroform** | 1.498 | 26.5 | 4.8 |
| **Dimethylformamide (DMF)** | 0.994 | 37.1 | 38.3 |
| **Ethanol** | 0.789 | 21.9 | 24 |
| **Water** | 1.000 | 72.8 | 80 |

However, EHD processes require low-viscosity and low-surface tension fluids, therefore the positive ER effect must be avoided, and the negative ER effect may be useful (**Table 1**); moreover, the EHD inks also require an appreciable dielectric constant (**Table 2**). The required low-viscosity values of the EHD inks highlight one of the major limitations of the electrorheological cells commercially available on the market for rotational rheometers, which were conceived for larger viscosity values. To promote the connection between the EHD processes and electrorheometry, Rijo and Galindo-Rosales have recently focused on electrorheological characterization of inks composed by 2D nanomaterials dispersed in a leaky dielectric viscoelastic fluid [39, 40], showing evidences that the electrical conductivity and concentration of nanomaterials affect the ink's rheology when an external electric field is applied. Additionally, the relative orientation between the electric and the flow fields is an important aspect to consider. Typically, the electrorheological characterization is performed in a rotational rheometer equipped with an electrorheological cell, where the external electric field is oriented perpendicularly to the flow direction; however, that configuration does not mimic the actual conditions of any EHD printing technique at any location (**Figure 3-top**). The electrorheological cell developed by Sadek *et al.* for performing capillary breakup extensional electrorheometry (CaBEER) [41] allows to impose an electric field parallel to the direction of the flow, which mimics very well the configuration occurring during the filament thinning process in the EHD-jet printing process and electrospinning process (**Figure 3-bottom**).

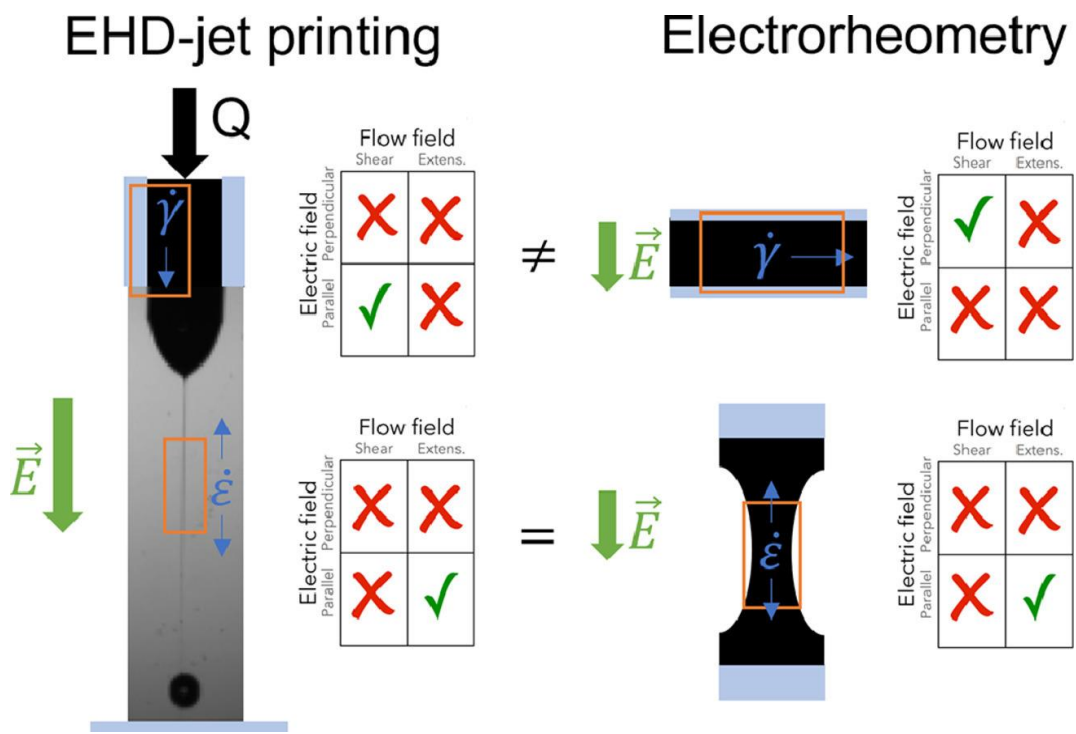

**Figure 3.** Electric and flow field configurations occurring in an EHD printing process compared with the configurations allowed by the electrorheological cells developed so far for rotational and extensional rheometers. Reproduced with permission from [39].

This work aims to define the limitations of current electrorheological cells, propose methods to mitigate their impact on measurements to avoid inaccurate data collection, and show how microfluidics can offer innovative solutions to overcome these technical limitations on a macroscale.

2. Measurements of the Dielectric Properties

To perform a proper electrorheological characterization, it is very important to know in advance the dielectric properties of fluids. The interfacial polarization, also known as Maxwell-Wagner polarization, is the main mechanism responsible for the ER effect of suspensions [42]. According to Block *et al*. [43] the relaxation frequency is a key parameter to observe the ER effect and it is proportional to the polarization rate. The relaxation frequency occurs when the dielectric loss factor ($\varepsilon''$) reaches the maximum value [43]. For fluids with a strong ER effect, this frequency must be in the range of $10^2$ and $10^5$ Hz, as schematically represented in **Figure 4**.

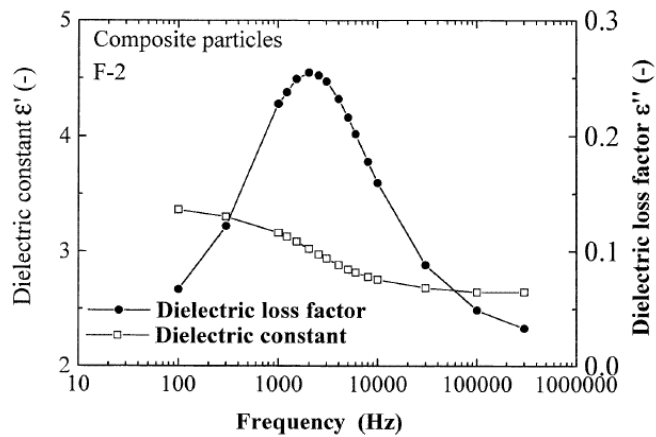

**Figure 4.** Dielectric constant ($\varepsilon'$) and loss factor ($\varepsilon''$) of a fluid studied by Ikazaki *et al*. [44] Reprinted with permission of [44].

Ikazaki *et al*. [44] reported that the polarization rate is responsible for maintaining the chain-like structures formed by particle dispersed in a carrier fluid under an applied electric field. The formation and destruction of the chain-like structures depends on the ability of the polarization rate to overcome the shear rate applied to the fluid. Moreover, the chain-like structures cannot be formed when the polarization rate is too low, but also, if it is too high, the structures are formed, and the particles are easily rotated by the motion imposed by the shear rate and the induced dipoles formed in the particles will realign according to the electric field direction [44]. This realignment will promote the repulsive forces between particles and not the attractive forces to sustain the particle structures. Further, the difference in dielectric constants between $10^2$ Hz and $10^5$ Hz ($\Delta\varepsilon' = \varepsilon'_{10^2 Hz} - \varepsilon'_{10^5 Hz}$) is also important to evaluate the ER effect. If the relaxation frequency is in the range of $10^2 – 10^5$ Hz, the higher the difference of dielectric constant, the greater the ER effect will be.

There are two methods to measure the dielectric properties of fluids. The first method consists of using a Broadband Dielectric Spectrometer at 1 V with an alternating current frequency between 1 Hz and 200 kHz [42]. The liquid is confined in a special cell consisting of two brass electrodes with a 0.5 mm thick Teflon spacer in between. The second method is a homemade solution proposed by Rijo and Galindo-Rosales [39, 40] and it consists of using a LCR meter (Keysight E4980AL) that applies an alternating current window between 20 Hz and 300 kHz at 1 V to the liquid confined in two horizontal circular stainless-steel plates with 50 mm diameter and a gap between plates of 0.5 mm (**Figure 5a**). This latter method can be implemented with relative ease in a rotational rheometer equipped with an electrorheological cell.

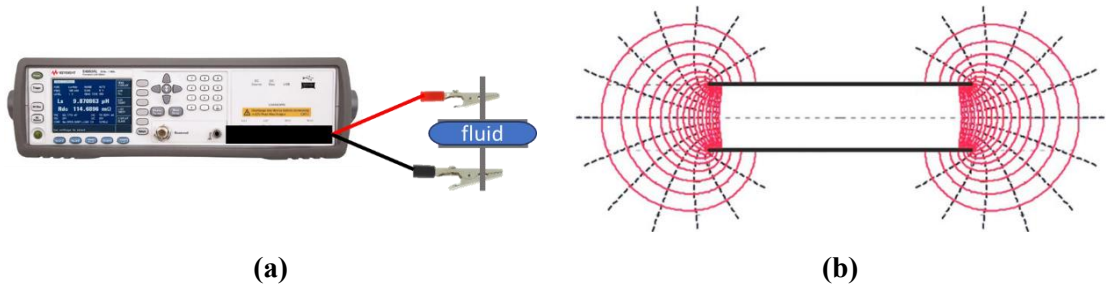

**(a)**                                        **(b)**

**Figure 5. (a)** Schematic illustration of the experimental setup used to measure the dielectric properties proposed by Rijo and Galindo-Rosales. **(b)** Parallel-plate capacitor with the fringing effect (curved red lines). Reprinted with permission of [45]

The system used by Rijo and Galindo-Rosales [39, 40] to determine the dielectric properties based on capacitance is very similar to a parallel-plate capacitor. The dielectric constant (ε') and dielectric loss factor (ε'') are calculated as follows [46]:

$$\varepsilon' = C_{fluid}/C_0 \qquad (3)$$
$$C_0 = \varepsilon_0 A/h \qquad (4)$$
$$\varepsilon'' = \varepsilon' \tan(\delta) \qquad (5)$$

where $C_{fluid}$ and $C_0$ are the capacitance of the fluid tested and air, respectively, $A$ is the surface area of the plate, $h$ is the distance between the plates, and $\tan(\delta)$ is the loss factor. There are three factors that influence the value of the capacitance [47]: (i) the surface area of the plate, (ii) the space between the plates and (iii) the permittivity of the material. The capacitance's value is strongly influenced by the fringing effect when parallel-plate capacitor does not have an appropriate ratio between the length of the plate, i.e., radius of the plate, and the gap between plates [48]. The fringing effects occur at the edges of the electrodes where the electric field becomes non-uniform and extends out into the surrounding space (**Figure 5b**).

According to Xu *et al*. [48], the fringing effects can be neglected when the ratio between the size of the electrode and the gap between electrodes is higher than 10. When this ratio is lower than 10, the fringing effects increase the capacitance's value of the fluid and the electrostatic force [49]. Considering this effect, Rijo and Galindo-Rosales used electrodes with 50 mm diameter and a gap of 0.5 mm to ensure a ratio of 50 (> 10) to avoid the fringing effects during the dielectric properties' measurements.

## 3. Shear ElectroRheology

The rotational rheometers equipped with electrorheological (ER) cells have been a key approach to characterize the rheological properties of fluids when an electric field is applied. However, two different ways to create an electric field are commercially available. The MCR 3xx and 5xx rheometers (Anton Paar) and ARES G2 rheometer (TA Instrument) use the same approach to produce an electric field (**Figure 6a** and **b**), *i.e.* the top geometry is in direct contact with an electrified wire connected to an external high voltage power supply. That wire is responsible for additional friction, resulting in artificially higher viscosity values if not considered during the calibration procedure. The rigidity of this wire is affected by Joules' effect, which is directly proportional to the current circulating through the fluid sample. Therefore, the higher the conductivity of the fluid sample under the same electric field results in a higher Joule effect, reducing the rigidity of the wire. Moreover, higher Joule effects may jeopardize the temperature stability of the sample.

In contrast, the ER cell of the former Bohlin Gemini CVOR 150 uses an electrolyte solution which is responsible for supplying the voltage generated by the external high-voltage source to the geometry (**Figure 6c**).

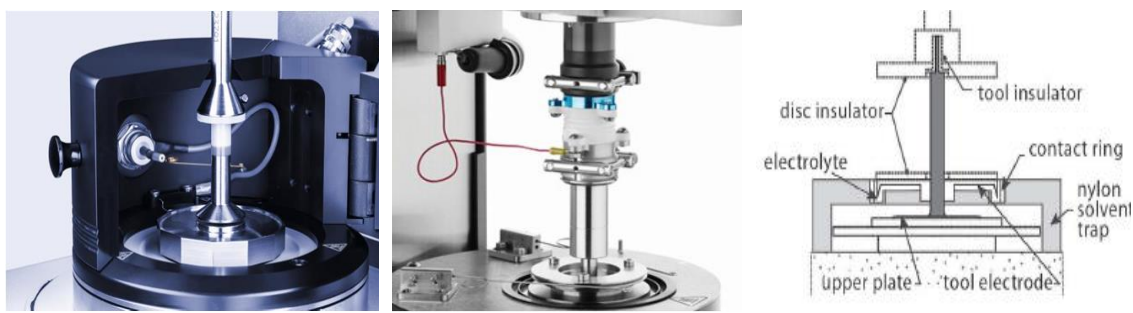

**(a)** **(b)** **(c)**

**Figure 6.** Electrorheogical cells developed by: **(a)** Anton Paar, **(b)** TA Instruments and **(c)** Bohlin Gemini. Reprinted with permissions of [50].

This is an alternative way to electrify the top geometry avoids the extra friction effect due to the wire. Peer *et al.* [50] compared the electrorheological measurements of polyaniline powder dispersed in silicone oil obtained from the electrorheological cells present in **Figure 6 a** and **c**. The authors concluded that the rheological data obtained by Anton Paar equipped with its wired-electrorheological cell provides higher values of shear viscosity, storage and loss moduli than those obtained by Bohlin Gemini electrolyte-electrorheological cell; moreover, they also noticed that this increment was geometry dependent, as it was more significant for the parallel-plates (PP) than for the concentric cylinders (CC), as depicted in **Figure 7**.

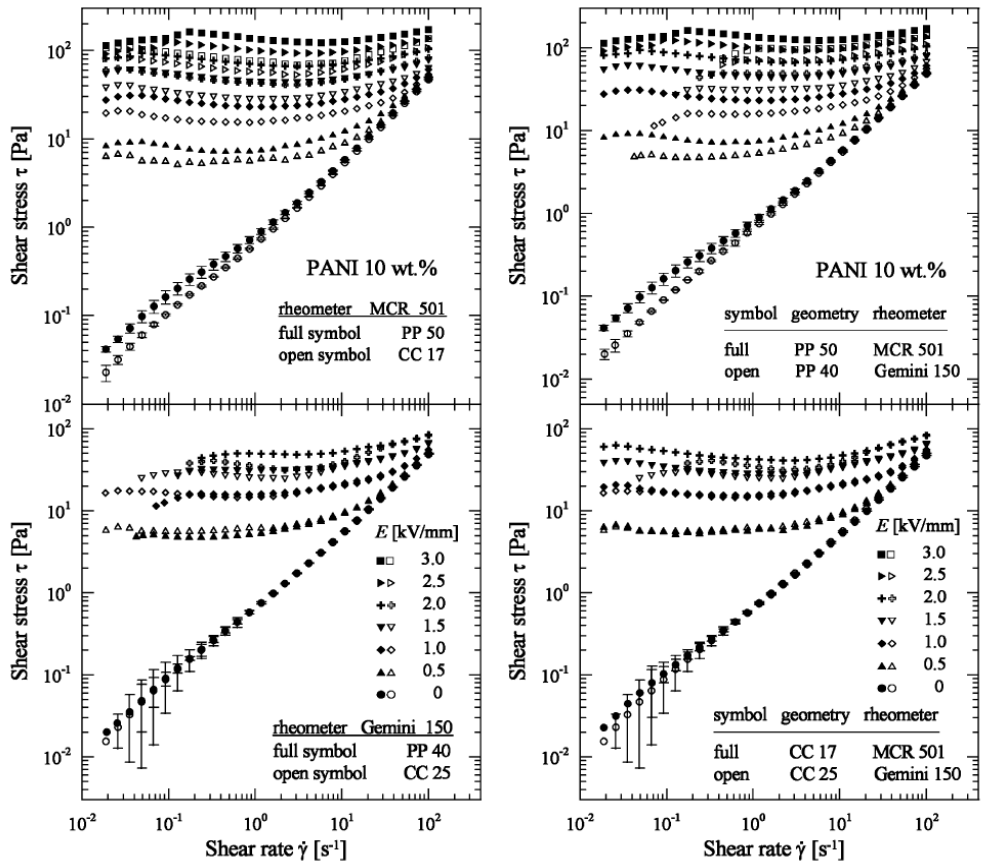

**Figure 7.** Shear stress dependent on shear rate: **(left)** different geometrical arrangements (PP and CC, same rheometer) and **(right)** same geometrical arrangement (PP or CC, different rheometers). Reprinted with permission of [50].

Moreover, Peer *et al.* [50] also found that the additional friction effect can be negligible when the suspensions have a viscous carrier fluid, such as silicone oil, or a stronger electrorheological effect. However, these considerations are not valid when the suspensions have a weak electrorheological effect, or the carrier fluid has viscosity less than 1 Pa·s. **Figure 8** shows the shear viscosity measurements of ethyl cellulose (EC) dissolved in toluene for different concentrations of polymer. The presence of the wire in the system influences the viscosity curves regardless of the polymer concentration. Since toluene is a Newtonian fluid, the experimental results obtained can lead to erroneous interpretations of the fluid's behavior, i.e., assuming the presence of shear-thinning in a Newtonian fluid at low and moderate shear rates (**Figure 8**). The same happens when low-viscosity polymer solutions are used, i.e., the observation of two shear-thinning slopes separated by a plateau of constant viscosity values. These results diverge from the viscosity curves obtained when the fluids are tested under normal conditions for the same range of shear rates, i.e. without the presence of wire in the mechanical system (**Figure 8**). Even when applying the correction mechanism present in RheoCompass, we propose establishing a new minimum torque value to remove the bad data present for low and moderate shear rates to account for this friction effect. This new torque value is around 30 µN·m instead of the 0.1 µN·m defined by the Anton Paar for the MCR 301 rheometer used in this work.

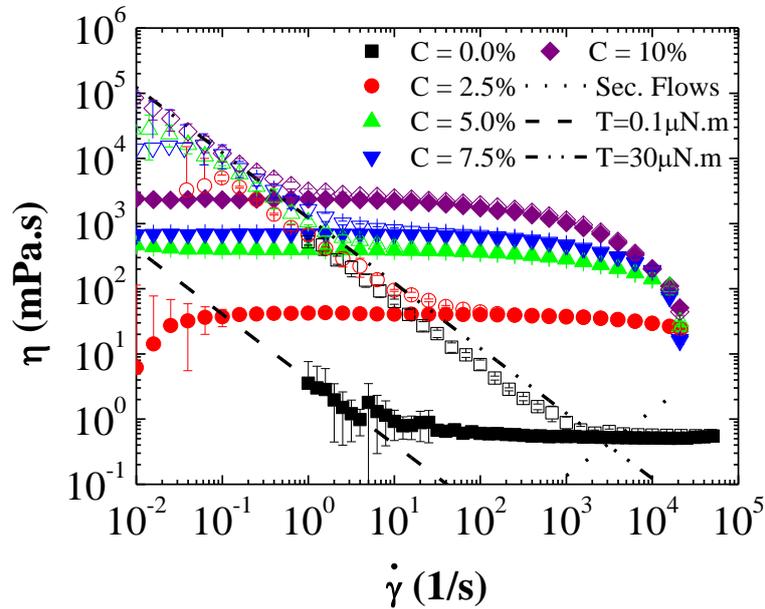

**Figure 8.** Shear viscosity dependent on shear stress for different concentrations of ethyl cellulose (EC) dissolved in toluene without application of electric field. Fill and open symbols represent the experimental data obtained from a normal rheological cell (T = 0.1 µN.m) and ER cell (T = 30 µN.m) developed by Anton Paar.

Considering the RheoCompass' correction mechanism and the new minimum torque value, Rijo and Galindo-Rosales [39, 40] were able to correctly analyze the electrorheological properties of weakly viscoelastic fluids (**Figure 9a**); however, Roman *et al.* [31] did not have this limitation when studying the electrorheological properties of suspension with low-viscosity carrier fluids, thanks to the electrorheological cell design using an electrolyte for applying the voltage to the upper plate (**Figure 9b**). The suspension studied by Roman *et al.* [31] consisted of 1 wt.% of polyaniline-graphene particles dispersed in silicone oil with a kinematic viscosity of 20 cSt and the suspension studied by Rijo and Galindo-Rosales [40] was 0.25 wt.% of molybdenum disulfide ($MoS_2$) dispersed in 2.5% w/v of ethyl cellulose dissolved in toluene. The test conditions used in the work of Roman *et al.* [31] were a parallel-plate geometry of 40 mm diameter with a gap of 0.5 mm and a parallel-plate geometry of 50 mm diameter with a gap of 0.10 mm was used by Rijo and Galindo-Rosales [40]. The minimum shear stress (τ) as follows [19]:

$$\tau_{min} = \frac{3T_{min}}{2\pi R^3}, \qquad (1)$$

where $T_{min}$ and $R$ are the minimum torque and radius of the plate, respectively. Assuming a linear velocity profile, the minimum viscosity value will be given by Equation 2:

$$\mu_{min} = \frac{2H}{\pi R^4}\frac{T_{min}}{\Omega}. \qquad (2)$$

where $H$ and $\Omega$ are the gap between the plates and the rotational speed, respectively. According to Equation 2, for a given fluid, having an electrorheological cell with a higher $T_{min}$, that is the one with the wire, will require the use of larger diameter plate and higher shear rates ($\dot{\gamma} = \frac{\Omega R}{H}$) to allow reliable measurements. Considering these limitations, it seems that the electrorheological cell developed by Bohlin Gemini is more suitable for studying the behavior of low-viscosity ER fluids at shear rates below 100 s$^{-1}$, while the ER cell developed by Anton Paar is more suitable for high shear rates or viscous fluids.

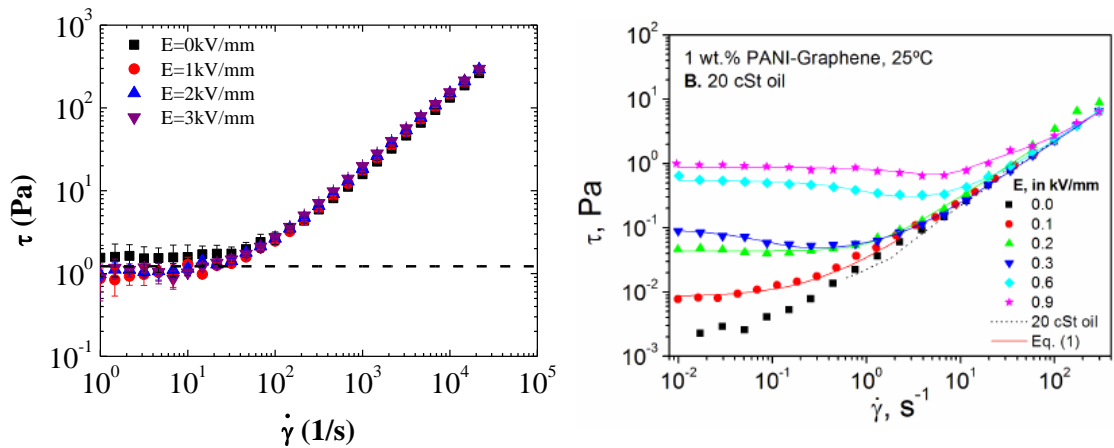

**(a)** **(b)**

**Figure 9.** Qualitative comparison of the flow curves obtained by means of an electrorheological cell using a wire **(a)** and an electrolyte solution **(b)**. Reprinted with permissions of [31].

The second limitation when using electrorheological cells is related to the maximum electrical current allowed by the voltage generator. In the experimental setup used by García-Morales *et al.* [42], the power source used in the ARES-G2 rheometer allows the generation of electric fields with a maximum current intensity of 20 mA, which is around 20 times higher than the maximum value allowed by the voltage generator used in the MCR 301 rheometer used in the experimental setup of Rijo and Galindo-Rosales [39, 40]. This limitation means that the electrorheological characterization of fluids is limited by electrical conductivity of the fluid sample. This makes it very difficult or practically impossible to characterize fluids formulated with slightly polar (ethanol) or polar (water) solvents, and with non-polar solvents laden with electrically conductive particles (graphene) or polymers (polyaniline). This is a major limitation that makes it difficult to rheologically characterize the effect of the electric field on the inks typically used in electrohydrodynamic processes. To overcome this limitation, we propose the following solution. Regardless of its physical state, we assume that the working fluid respects Ohm's law ($U = R_{fluid}.I$) when a voltage ($U$) and current ($I$) are applied to the system [47]. The electrical resistance of the fluid ($R_{fluid}$) filling the gap of a parallel plate geometry can be calculated as follows [47]:

$$R_{fluid} = \frac{\rho_e H}{A}, \qquad (3)$$

where $H$ and $A$ are the gap between the plates and area of the plate, respectively, and $\rho_e$ is the electrical resistivity of the fluid which is the inverse of the electrical conductivity ($\rho_e = 1/\sigma_e$). Combining the Ohm's law and Equation 3, it is possible to determine the maximum conductivity that can be measured for a given voltage generator (Equation 4):

$$\sigma_{e,max} = \frac{H}{\pi R^2}\frac{I}{U} = \frac{1}{\pi R^2}\frac{I_{max}}{E}, \qquad (4)$$

where $E = \frac{U}{H}$ is the electric field imposed to the fluid and $I_{max}$ is the maximum electrical current allowed by the voltage generator. **Figure 10** shows the relative importance of the diameter of the plate, the imposed electric field and the maximum intensity of the voltage generator in order to allow the electrorheological characterization of common solvents used in EHD printing process. It becomes evident that having a voltage generator allowing higher current intensity values is paramount for characterizing higher conductivity fluids. **Figure 11** shows the best option for the maximum current intensity of a high-voltage power supply to be coupled to the rotational rheometers. For a current intensity of 200 mA, it would be possible to carry out the electrorheological characterization of slightly polar solvents used in electrohydrodynamic techniques.

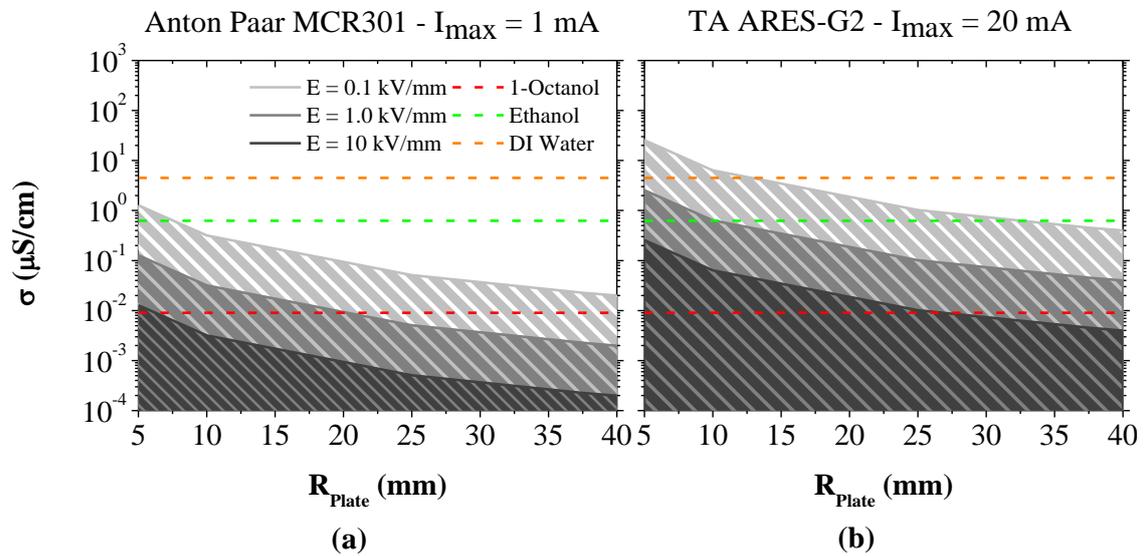

**Figure 10.** Comparison of the maximum conductivity allowed for each electric field and plate radius for the voltage generator commercialized by Anton Paar **(a)** and TA Instruments **(b)**.

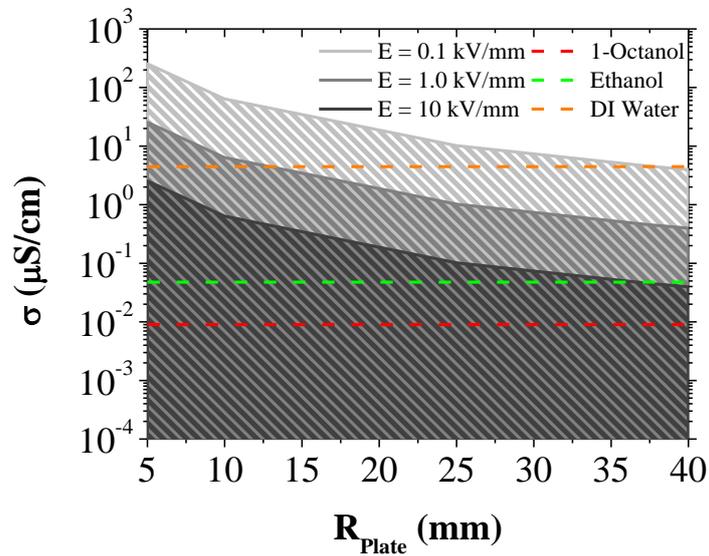

**Figure 11.** The dependence of the electrical conductivity on the plate radius of a rotational rheometer coupled hypothetically to a high-voltage power supply that allows a maximum current intensity of 200 mA.

## 4. Extensional ElectroRheology

Many manufacturing processes, including fiber spinning, blow molding, spray coating, electrospinning, electrospray and EHD-jet printing [4, 51, 52], are governed by elongational deformations. Hence, understanding the extensional properties of the fluids used in these processes is of paramount importance. In an effort to understand the effect of the electric field on extensional rheological properties, Sadek *et al.* [41] pioneered the development of an electrorheological cell that can be used in the capillary break-up extensional rheometer (CaBER). The authors reported that the extensional behavior of the electrorheological fluids consisting of a dispersion of cornstarch in olive oil depends on the particle concentration and the applied voltage; moreover, they confirmed that the Hencky strain is a relevant to characterize the ER fluids behavior [41]. García-Ortiz *et al.* [53] further developed that study analyzing the influence of the polarity of the electric field during the filament thinning process for Newtonian and ER fluids. They observed that the filament lives longer when polarity is against gravity. Later, Rubio *et al.* [54] measured the filament electrical conductivity and extensional relaxation time for polyethylene oxide dissolved in deionized water and in mixture of glycerol and deionized water. The authors reported some limitations during the experimental campaign. The one was the high-voltage power supply could not ensure a constant electric field strength due to the electrical conductivity of the fluids. To overcome this limitation, the authors proposed to use a system of resistors in series between the fluid and the high-voltage power supply [54], as shown schematically in **Figure 12.** However, this is not the best option to solve the problem, because the true voltage applied to the fluid is a little lower than the voltage applies to the whole system.

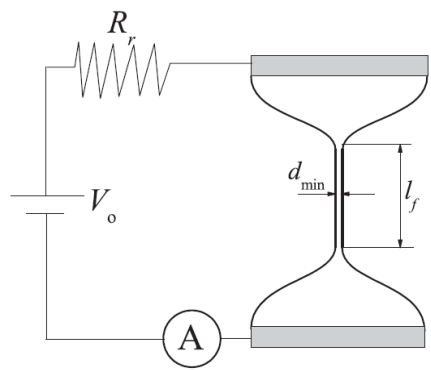

**Figure 12.** Schematic solution proposed by Rubio *et al.* [54]. Reprinted with permission of [54].

The diameters of the plates typically used in CaBER are 4, 6 and 8 mm. Applying the approach followed in section 2 to determine the maximum current value that allows us to characterize the electrorheological properties of slightly polar solvents under extension flow, **Figure 13** shows that a voltage source with a maximum current intensity of 20 mA is sufficient to overcome the limitations present when a voltage source of 1 mA is used. In addition, the analysis shown in **Figure 13** for a 20 mA voltage source makes it possible to study the electrorheological properties of DI water for a range of electric field strengths between 0.1 kV/mm and 5 kV/mm for a 4 mm diameter plate without the need to manufacture a smaller plate for the CaBER. This analysis is consistent with the experimental work done by Rubio *et al.* [54], where the authors were able to study the electrical and rheological properties of aqueous polyethylene oxide solutions when subjected to strain stresses using a CaBER coupled to a high-voltage power supply of 20 mA. However, associated heat and mass transfer problems would be expected, particularly at the later stages of the thinning process, when the cross-sectional area of the filament is minimal.

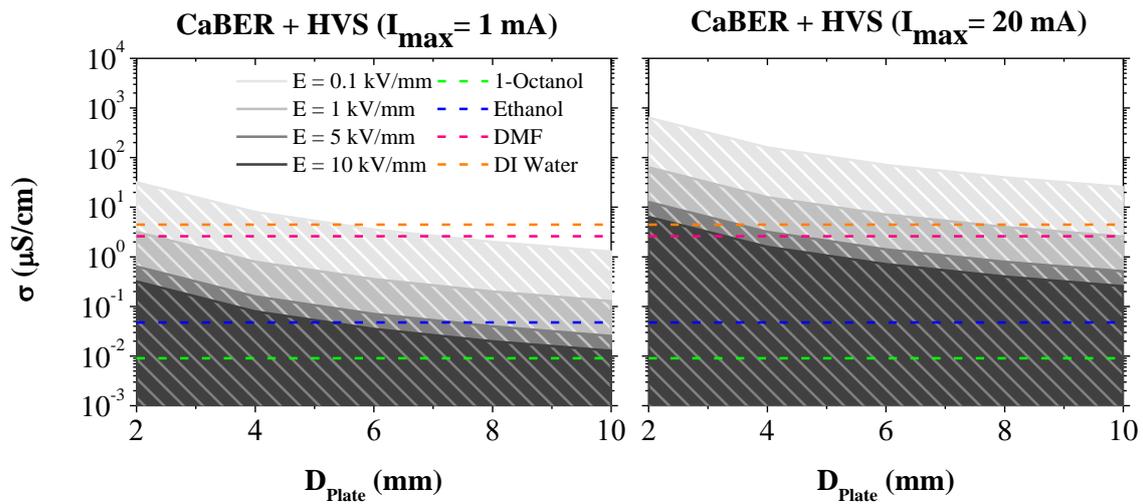

**Figure 13.** The dependence of the electrical conductivity on the plate diameter of a capillary breakup extensional rheometer (CaBER) coupled to a high-voltage power supply (HVS) of 1 mA **(left)** and coupled hypothetically to HVS of 20 mA **(right)**.

Although this limitation, Rubio *et al.* [54] could study the influence of the electrical conductivity with the minimum filament diameter during the thinning process. However, they could not ensure if the electrical conductivity depends on the electric field strength or there were other effects that could be present, such as Joule's effect [55]. To see this effect, two different experimental methods can be used: (i) use a color camera coupled with thermal lenses, and (ii) use the Schlieren technique [56]. In the first option, it is possible to observe the temperature gradient inside the filament during the thinning process. The second option consists of visualizing density variations in transparent media [56]. If the Joule's effect is present, the Schlieren images allows us to see the density variations promoted by the fluid's evaporation. Nevertheless, these techniques would require further experiments to be tested/validated either using CaBEER.

In last decade, Dinic *et al.* [57] proposed a new experimental method, known as dripping-onto-substrate (DOS) extensional rheometry, for determining the relaxation time and the extensional viscosity for low-viscosity elastic liquids, beyond the range measurable in the standard geometries used in the CaBER device. This methodology may, in principle, face the same problems as CaBER regarding partial evaporation; however, Robertson and Calabrese [58] successfully developed a chamber to enclose the sample in an environment saturated with solvent vapor in order to mitigate evaporation during dripping-onto-substrate (DoS) extensional rheology measurements. Rubio *et al*. [59] implemented an electrified version of this DoS methodology (**Figure 14**) to study experimentally and numerically the thinning of Newtonian leaky-dielectric filaments subjected to an axial electric field; although they considered moderately viscous liquids with high electrical permittivity, which are far from EHD printing conditions, they confirmed that the electric force delays the free surface pinching due to polarization stress. In their work, the partial evaporation of the solvent was not an issue, and the effect of viscoelasticity was not considered.

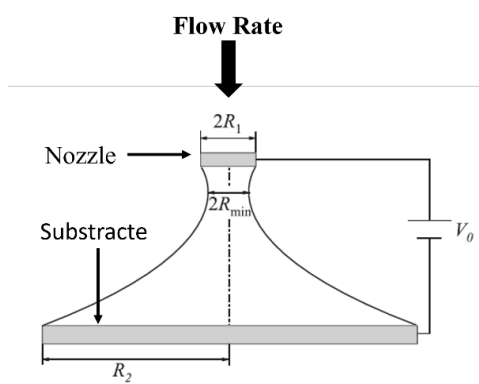

**Figure 14.** Electrified version of the Drop-On-Substrate methodology for performing electrorheology. Reprinted from [59-62].

Rijo and Galindo-Rosales [39, 40] studied the extensional rheological properties of nonpolar viscoelastic inks laden with 2D nanoparticles in the CaBEER. They noted the need to use the slow retraction method (SRM) developed by Campo-Deaño and Clasen [63] to minimize the inertial effects present in the fluid during the thinning process. Despite the high volatility of the solvent, partial evaporation of the fluid during the experiment was discarded due to the low electrical conductivity and the short experiment time. However, Rijo and Galindo-Rosales [39, 40] observed that the 2D nanoparticle suffered from migration and the ethyl cellulose providing elasticity to the carrier fluid induced the formation of vortices inside the liquid bridge under the influence of an electric field. The 2D nanoparticle migration affects the stability and homogeneity of the fluid under study during the experiment timescale. Moreover, the presence of recirculations may jeopardize the condition of uniaxial elongation flow due to the presence of local shear rates, limiting the usefulness of the study to a qualitative comparison of the apparent relaxation time. However, it has been confirmed that the CaBEER allows to replicate very well the conditions occurring in the EHD printing process, where the consistency and homogeneity of the ink formulated with colloidal particles is "disrupted during the ink jet formation and regions with low and high concentrations are created" [64]; additionally tangential electrical stresses acting on the liquid-gas interface in the CaBEER experiments induce the vortex formation, as in the Taylor's cone during the EHD printing processes [65]. Furthermore, the shape of end-drops at the end of the CaBEER experiments allows to assess if a formulation will be able to form the Taylor's cone in EHD printing process [40, 66].

## 5. Microfluidics

Microfluidics is the science and technology of systems that process or manipulate very small amounts of fluids in geometries with characteristic length scales below one millimeter [67]. The small length scales in microfluidics enable the generation of flows with high deformation rates while maintaining a low Reynolds number ($Re$). These unique flow characteristics promote strong viscoelastic effects, quantified by the elasticity number ($El$), which scales inversely with the square of the characteristic length ($L$) [68]. As the scale is reduced, these effects are enhanced. These distinctive flow properties, combined with numerical optimization techniques [69-74] provide a rich platform for rheologists to conduct rheometric investigations of non-Newtonian fluid flow phenomena at small scales. This approach overcomes the limitations of macroscopic rheometry [75-78] and allows exploration of regions in the $Wi$–$Re$ map unreachable by commercial rheometers [79].

Additionally, a microfluidic-based rheometer-on-a-chip offers several practical advantages, such as low volume samples, minimize partial evaporation of solvents and the ability to serve as an online rheological sensor in various industrial processes. The VROC® Technology [80], commercialized by Rheosense [81], is currently the only device allowing the measurement of steady shear flow curves and can even be used to characterize complex fluids under extensional flows [82]. Until 2023, Formulaction commercialized the FluidicamRheo™, a shear microrheometer based on the technology developed by A. Colin and co-workers [83]. This device is an optical microfluidic viscometer where viscosity determination does not rely on pressure transducers [84]. Other approaches and designs of microrheometers exist [85, 86], but none of them have reached the market yet nor can perform electrorheology. Therefore, to the best of authors' knowledge, it is currently not possible to perform shear electrorheometry at the microscale.

The difficulty in providing homogeneous electric field throughout the whole volume of the fluid sample might have discouraged the exploration of electrorheometry at microscale. Recently, Galindo-Rosales *et al.* [87] disclosed an invention to perform microelectrorheometry, either shear or extensional. As depicted in **Figure 15**, there is a main microchannel through which the fluid sample flows, driven by either a syringe or a pressure pump, and two auxiliary microchannels, filled with a metal or a metal alloy, that are parallel to each other, either arranged parallel or perpendicular to the main microchannel. These auxiliary microchannels operate as electrodes for generating an electrical field to be applied to the fluid sample. With this configuration, it is possible to impose an external electric field parallel or perpendicular to the fluid flow, which, depending on the chosen geometry, can be designed for shear or extensional flow. This proposal ensures the homogeneity of the electric field within the zone of analysis, and it would be compatible with the VROC and FluidicamRheo devices; however, they have not been yet tested experimentally. Electrophoresis and electroosmosis can be foreseen as potential limitations to this device, requiring further research to select the right materials for their constructions [88].

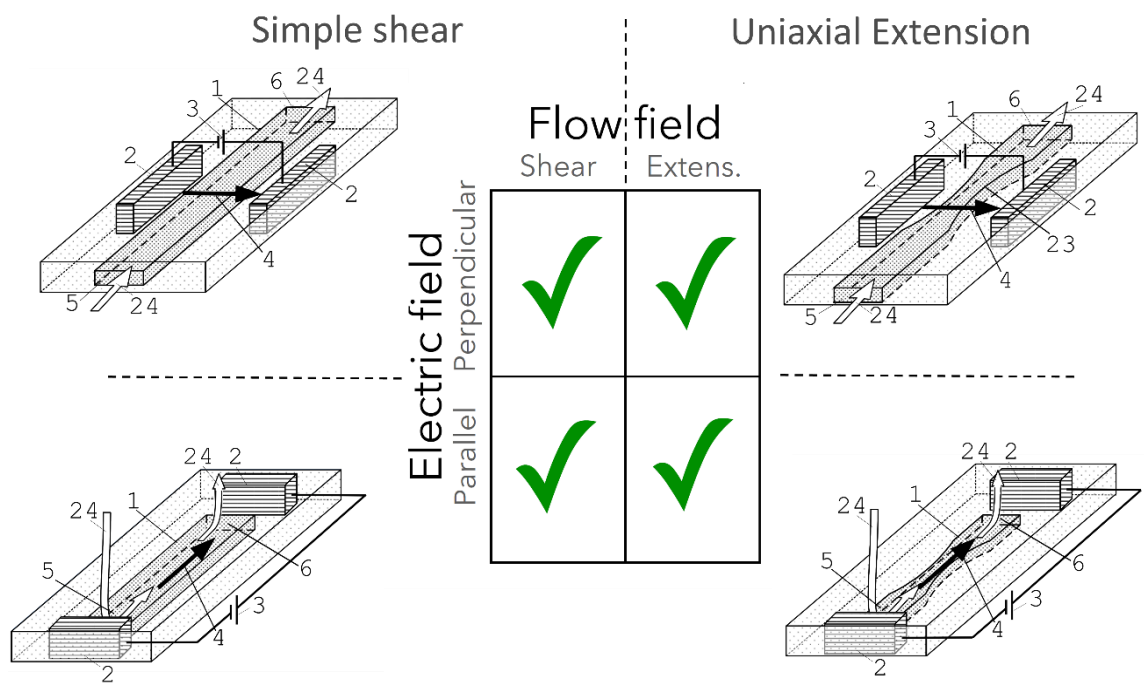

**Figure 15.** The four possible configurations for the microelectrorheomter allowing to expand the limitations of the current electrorheological cells available for the commercial rheometers. different configurations for the integrated Microelectrorheometer, all of them containing the following components: the main microchannel (1) through which the fluid sample flows in the direction indicated by (24), the external electric field (4) between the auxiliary microchannels (2) acting as electrodes by means of the voltage supplier (3). Reproduced from [87].

## 6. Final remarks

The electrical conductivity of the fluids used in EHD processes leads often relegates electrorheology to a secondary role. This lack of integration means that it is challenging to predict the effects of viscosity and relaxation time in the presence of an electric field when the fluid undergoes elongational or shear deformation. Knowing these fluid properties beforehand can help establish relationships between operating and fluid parameters, preventing clogging phenomena and optimizing EHD processes. For example, understanding beads-on-a-string formation can indicate whether nanofibers from electrospinning replicate this effect, or if satellite droplets would be generated during main droplet detachment in e-jet printing.

To reinforce the relationship between electrorheology and electrohydrodynamic, the authors suggest using the most suitable plate diameter and high-voltage power supply for shear rheological characterization of moderately conductive fluids, ensuring the highest benefit-cost ratio. For studying the relaxation time and the formation of beads-on-a-string, 4 mm plates and a 20 mA high-voltage power supply are the most suitable tools for CaBER.

A detailed knowledge of the dielectric properties is extremely important in electrorheology and EHD processes. The experimental setup proposed by Rijo and Galindo-Rosales offers an economical and effective solution, where the fringing effects can be neglected due to the high ratio of the radius of the plates and the distance between them.

## Acknowledgement

This work was financially supported by UI/BD/150886/2021 (FCT), LA/P/0045/2020 (ALiCE), UIDB/00532/2020, and UIDP/00532/2020 (CEFT), funded by national funds through FCT/MCTES (PIDDAC), and the program Stimulus of Scientific Employment, Individual Support-2020.03203.CEECIND.